\documentclass[cameraready]{Interspeech}

\title{Does Translation-Enhanced Speech Encoder Pre-training Affect Speech LLMs?}

\author[affiliation={1}]{Tomoya}{Mizumoto}
\author[affiliation={1}]{Yusuke}{Fujita}

\address{
    $^1$ SB Intuitions 
}

\email{tomoya.mizumoto@sbintuitions.co.jp, yusuke.fujita@sbintuitions.co.jp}


\usepackage{comment}
\usepackage[table]{xcolor}
\definecolor{lightgray}{gray}{0.8}

\begin{document}

\maketitle

\begin{abstract}
 Connecting a pre-trained speech encoder to a Large Language Model (LLM) is the standard architecture for building Speech LLMs.
However, a structural misalignment exists between the encoder and the LLM.
Unlike encoders based on automatic speech recognition, which often produce representations in separate language-specific spaces, LLMs operate within a unified language-agnostic space.
A mechanism is required to align the encoder's language-specific representations with the LLM's shared space.
We argue that speech translation provides a principled way to achieve this.
Unlike monolingual transcription, translation requires the model to bridge different languages and learn language-agnostic representations.
We experimentally evaluate the impact of incorporating translation objectives into speech encoder pre-training.
Our results demonstrate that translation-enhanced pre-training improves cross-modal integration and leads to superior performance across downstream Speech LLM tasks.
\end{abstract}

\section{Introduction}
To build Speech Large Language Models (Speech LLMs) capable of processing audio directly, a prevalent architecture integrates a pre-trained speech encoder with an LLM via a trainable adaptor~\cite{fathullah2023prompting,rubenstein2023audiopalm,hono-etal-2024-integrating,zhang2023speechgpt}. 
Within this framework, an adaptor maps continuous acoustic features into the embedding space of the LLM, facilitating seamless cross-modal understanding.
Unlike conventional cascaded systems requiring an intermediate text-transcription step, this direct mapping avoids compounding errors and allows the LLM to leverage denser, continuous representations.
To extract these representations, the prevailing practice is to initialize the speech encoder with robust models pre-trained primarily on self-supervised learning (SSL) or transcription-centric tasks, such as automatic speech recognition (ASR)~\cite{10.1109/TASLP.2021.3122291,comfomer,schneider19_interspeech}.

However, directly projecting these continuous embeddings into an LLM remains challenging due to the representational mismatch between speech and language embedding spaces~\cite{wang2023asru}.
The representation space of a text-based LLM is highly structured by deep semantics and abstract concepts, whereas that of SSL pre-trained audio encoders is organized according to acoustic and phonetic similarities~\cite{choi2024_interspeech}.
Although ASR pre-trained encoders can capture semantic information, their speech--text mappings are typically language-specific.
When such language-specific continuous embeddings are directly fed into an LLM, the lightweight adaptor must bridge the substantial gap between the phonetically driven acoustic features and the abstract semantic space of the LLM.
To resolve the root cause of this structural misalignment, we argue that the continuous representations of the speech encoder must be abstracted beyond language-specific boundaries.
To explicitly achieve a unified cross-lingual abstraction, the encoder must be trained on a task that demands language-agnostic semantic understanding.
By forcing the model to bridge different languages, the encoder can generate representations that naturally align with the LLM's universal semantic space.

Cross-lingual translation is a task that fundamentally meets this requirement. 
Unlike transcription, translating speech into a different language necessitates extracting the underlying meaning because the input audio and target text share no phonetic overlap.
Indeed, this objective has been incorporated into large-scale models such as Whisper~\cite{Radford2023whisper}.
Whisper now serves as a backbone encoder for numerous Speech LLMs~\cite{chu2024qwen2audiotechnicalreport,xu2025qwen25omnitechnicalreport,tang2024salmonn,fang2025llamaomni,zhang-etal-2025-soundwave}.
Interestingly, however, Whisper is primarily trained to translate non-English speech into English text, while its English audio is trained exclusively on monolingual transcription. 
Consequently, when the input is English, the encoder may not fully exhibit its capacity for semantic abstraction, thereby limiting its effectiveness as a backbone for Speech LLMs.

To investigate this possibility, we explore the impact of incorporating English-to-any translation into the pre-training of speech encoders. 
We rigorously compare the performance of models trained with different training objectives: 1) transcription-only, 2) transcription and translation to English only (following the original Whisper training paradigm), and 3) transcription and bidirectional English translation.
We hypothesize that the third approach will lead to more effective cross-modal alignment, particularly for English inputs.
Through our experiments, we demonstrate that encoders trained with this bidirectional English translation objective significantly improve cross-modal alignment, leading to superior performance on downstream Speech LLM tasks.

\section{Related Works}
 Ma et al.~\cite{ma-etal-2025-cross} demonstrated that Whisper inherently maps multilingual speech into a shared semantic space. 
By employing Whisper as a standard encoder--decoder and simply fine-tuning its decoder, they achieved zero-shot cross-lingual transfer, confirming that translation objectives naturally induce language-agnostic representations.
Although the utility of such translation-based representations is widely recognized, current Speech LLMs primarily rely on unidirectional translation.
Many state-of-the-art models simply adopt the pre-trained Whisper encoder~\cite{hu-etal-2024-wavllm,chu2023qwenaudioadvancinguniversalaudio,fang2025llamaomni2llmbasedrealtimespoken}, which is limited to translating non-English speech into English ($X \rightarrow \text{en}$).
Furthermore, when recent works train custom speech encoders, they often either omit cross-lingual translation entirely (e.g., Qwen3-Omni~\cite{xu2025qwen3omnitechnicalreport}) or replicate this unidirectional approach (e.g., TTA~\cite{liu2026ttatranscribetranslatealignment}).

To shed light on this underexplored area, we investigate the effect of a bidirectional translation objective, explicitly incorporating the English-to-any ($\text{en} \rightarrow X$) direction. 
Although prior works like the Open Whisper-style Speech Model (OWSM) v4~\cite{owsm-v4} have explored bidirectional translation within standalone sequence-to-sequence (Seq2Seq) architectures, the potential of this approach to fundamentally bridge the modality gap in LLM integration remains unexplored.
We directly integrate this bidirectionally aligned encoder into a strictly frozen LLM to systematically evaluate its impact on diverse downstream tasks, such as speech translation (ST), intent classification, and emotion recognition.

\section{Research Design}

The primary objective of this study is to investigate how incorporating translation tasks during speech encoder pre-training affects the cross-modal integration of the encoder with an LLM.
To this end, we designed a controlled experimental framework.
Maintaining the same overall model architecture and adaptor training process, we systematically compared the impact of different encoder pre-training objectives on the final capabilities of the Speech LLM.

\subsection{Architecture of the Speech LLM}
To isolate the impact of the speech encoder, we adopted a standard, widely used Speech LLM architecture~\cite{hono-etal-2024-integrating,Lakomkin2024,ma2024embarrassingly}.
This architecture consists of three main components: a speech encoder, a trainable adaptor, and an LLM.

Given an input speech waveform, the speech encoder extracts a sequence of continuous acoustic representations.
Because these continuous representations do not naturally align with the text-based input space of the LLM, an adaptor is required to bridge this alignment gap.
The adaptor projects the encoder's output into a sequence of speech embeddings that match the dimensionality of the LLM's text embeddings.
Finally, these projected speech embeddings are fed directly into the LLM.
In our framework, the LLM is kept completely frozen. This ensures that any differences in downstream performance originate solely from the representations generated by the speech encoder.

\subsection{Encoder Pre-training Strategy}
To systematically compare the effects of different pre-training objectives, we trained the speech encoders on two distinct generative tasks: multilingual transcription and cross-lingual translation.
To effectively handle both tasks during the pre-training phase, we adopted a standard Seq2Seq architecture.
Specifically, we adopted the same model architecture as Whisper~\cite{Radford2023whisper}, as it is widely used as a speech encoder in numerous prior studies and Speech LLM systems.

During the pre-training phase, the acoustic encoder extracts continuous representations from the input audio, and the text decoder generates the corresponding target text.
Once this pre-training was complete, we discarded the decoder and extracted only the trained encoder.
This standalone encoder was then integrated into our Speech LLM pipeline, where it provides continuous speech representations to the LLM.

To evaluate our hypothesis under practical computational constraints, we focused our experiments on a subset of four languages. 
As replicating the massive training scale of the original Whisper model across dozens of languages is computationally prohibitive, we selected languages for which substantial amounts of training data are accessible: English, Japanese, Chinese, and German.
By restricting the language set, we ensured a sufficient volume of training data per language while maintaining a strictly controlled experimental environment.
Using this targeted subset, we established three distinct pre-training configurations:
\begin{itemize}
    \item \textbf{ASR-Only}: The model was trained exclusively on multilingual transcription across all four languages.
    This setup served as our baseline to observe whether transcription alone can form a unified semantic space or if the representations remain confined within language-specific boundaries.
    
    \item \textbf{ASR \& ST ($X \rightarrow \text{en}$)}: The model was trained on multilingual transcription alongside translation from the non-English languages (Japanese, Chinese, and German) into English. 
    This configuration inherently creates an asymmetric translation objective. We adopted this setup to mirror the training conditions of the original Whisper model, which strictly omits translation from English.
    
    \item \textbf{ASR \& ST ($X \leftrightarrow \text{en}$)}: In addition to multilingual transcription, the model was trained on translation between English and the other three languages. 
    Critically, this configuration included translation from English into Japanese, Chinese, and German, forcing the encoder to decouple meaning from surface linguistic forms for all input languages symmetrically.
\end{itemize}

\subsection{Prompt Formatting for Bidirectional Translation}
The original Whisper is limited to X $\rightarrow$ English translation, as its prompt format lacks a mechanism to specify the target language. 
To enable bidirectional translation between any pair of the selected languages, we followed the multi-target approach of OWSM~\cite{peng2023asru} and introduced a redesigned decoder prompt.

Our format explicitly differentiates between provided conditions and predicted attributes.
The target language token is placed before the task token as a pre-defined condition.
Conversely, since the source language must be estimated from the input audio, its token is positioned after the task token as a predicted property.
For example, translating German audio into English is structured as \verb+<|BOS|>+\verb+<|en|>+\verb+<|translate|>+\verb+<|de|>+, while a transcription task follows the sequence \verb+<|BOS|>+\verb+<|de|>+\verb+<|transcribe|>+\verb+<|de|>+. 
This modification explicitly conditions the task on the target language, enabling effective bidirectional generation.

\section{Experiments}
\begin{table*}[t]
\centering
\caption{Downstream task performance across ASR and ST tasks for 1B and 3B LLMs. For ST (en $\rightarrow$ X), scores are formatted as CoVoST2 / FLEURS. ST (X $\rightarrow$ en) is evaluated solely on FLEURS. "Seen in Pre-training" indicates target languages included during the foundational Seq2Seq pre-training phase, whereas "Unseen in Pre-training" targets were introduced only during the subsequent Speech LLM training phase.}
\label{tab:main_results}
\vspace{-4mm}
\resizebox{\textwidth}{!}{
\begin{tabular}{l cccc ccc ccc cccc}
\toprule
& \multicolumn{4}{c}{\textbf{ASR}} 
& \multicolumn{3}{c}{\textbf{ST} (X $\rightarrow$ en)} 
& \multicolumn{7}{c}{\textbf{ST} (en $\rightarrow$ X)} \\
\textbf{Pre-training} & \multicolumn{4}{c}{(WER/CER $\downarrow$)}
& \multicolumn{3}{c}{(BLEU $\uparrow$)}
& \multicolumn{3}{c}{\textbf{Seen in Pre-training} (BLEU $\uparrow$)}
& \multicolumn{4}{c}{\textbf{Unseen in Pre-training} (BLEU $\uparrow$)} \\
\cmidrule(lr){2-5} \cmidrule(lr){6-8} \cmidrule(lr){9-11} \cmidrule(lr){12-15}
\textbf{Task Mixture} & \textbf{en} & \textbf{ja} & \textbf{zh} & \textbf{de} 
& \textbf{ja} & \textbf{zh} & \textbf{de} 
& \textbf{ja} & \textbf{zh} & \textbf{de} & \textbf{fa} & \textbf{id} & \textbf{sv} & \textbf{tr} \\
\midrule
\multicolumn{15}{c}{\textbf{LLM: Llama-3.2-1B-Instruct}} \\
\midrule
ASR Only                           & 16.6 & 29.2 & 30.0 & 27.1 & 7.1 & 7.2 & 21.3 & 15.4/13.5 & 21.5/18.7 & 16.6/14.7 & 5.8/5.9 & 16.4/18.6 & 16.5/17.2 & 3.7/3.8 \\
ASR \& ST (X $\rightarrow$ en)     & 16.3 & 21.1 & 25.9 & 26.1 & 10.5 & 10.0 & 23.4 & 15.9/13.7 & 21.4/18.7 & 16.8/15.5 & 6.1/6.4 & 16.6/19.2 & 16.7/17.6 & 3.9/3.8 \\
ASR \& ST (X $\leftrightarrow$ en) & \bf{14.6} & \bf{19.7} & \bf{23.0} & \bf{24.3} & \bf{11.8} & \bf{11.3} & \bf{23.9} & \bf{18.2}/\bf{15.7} & \bf{24.8}/\bf{21.3} & \bf{19.3}/\bf{18.2} & \bf{7.7}/\bf{7.6} & \bf{19.4}/\bf{22.0} & \bf{19.9}/\bf{20.9} & \bf{5.3}/\bf{4.5} \\
\midrule
\multicolumn{15}{c}{\textbf{LLM: Llama-3.2-3B-Instruct}} \\
\midrule
ASR Only                           & 11.6 & 22.4 & 24.3 & 42.5 & 11.6 & 11.4 & 28.7 & 20.8/18.9 & 28.3/26.5 & 22.1/22.2 & 9.8/12.0 & 22.5/27.0 & 23.7/24.8 & 7.6/8.8 \\
ASR \& ST (X $\rightarrow$ en)     & 11.6 & 16.6 & 21.4 & 26.2 & \bf{15.1} & 14.5 & 30.0 & 20.9/19.6 & 28.6/26.9 & 22.4/22.3 & 10.1/12.2 & 22.7/26.5 & 23.8/25.6 & 7.8/9.0 \\
ASR \& ST (X $\leftrightarrow$ en) & \bf{11.0} & \bf{15.8} & \bf{21.1} & \bf{24.3} & \bf{15.1} & \bf{15.5} & \bf{30.9} & \bf{22.7}/\bf{21.3} & \bf{30.8}/\bf{28.5} & \bf{24.2}/\bf{24.5} & \bf{11.3}/\bf{13.2} & \bf{24.5}/\bf{28.5} & \bf{26.3}/\bf{27.2} & \bf{9.0}/\bf{10.6} \\
\bottomrule
\end{tabular}
}
\vspace{-4mm}

\end{table*}
In this section, we evaluate the impact of the different pre-training configurations by comparing the downstream generative performance of the fully integrated Speech LLM.

\subsection{Training Data and Experimental Setup}
\textbf{Base Model Pre-training Data:} We compiled a large-scale multilingual speech dataset comprising approximately 130k hours of audio across the four target languages: English (73.6k hours), Japanese (36.2k hours), German (10.0k hours), and Chinese (9.8k hours).
To ensure diverse acoustic and linguistic coverage, we constructed this dataset by aggregating several widely used public corpora.
Specifically, we utilized LibriSpeech~\cite{librispeech} for English, ReazonSpeech~\cite{reazonspeech} for Japanese, Multilingual LibriSpeech~\cite{Pratap2020MLSAL} for German, and WenetSpeech~\cite{zhang2022wenetspeech} for Chinese.
Furthermore, to augment the data volume and enhance bidirectional coverage, we incorporated subsets of YODAS-OWSMv4~\cite{owsm-v4} and Common Voice~\cite{commonvoice:2020} across all four languages.
While foundation models such as Whisper (680k hours) and OWSM v4 (320k hours) utilize significantly larger datasets across dozens of languages, focusing our computational resources on these four specific languages ensured that our 130k-hour corpus provided a sufficiently dense and robust training signal to evaluate our hypothesis.

Because these corpora primarily contain monolingual transcriptions, we synthetically constructed the parallel data required for our translation tasks.
Specifically, we employed an LLM, Qwen2.5-32B-Instruct~\cite{qwen2025qwen25technicalreport}, to generate translations from the original transcriptions into the respective target languages.
The task mixture ratios were strictly defined based on the training objective.
For the ASR-Only baseline, 100\% of the data across all languages was allocated to the transcription task.
For the ASR \& ST ($X \rightarrow \text{en}$) configuration, English data was used exclusively for transcription (100\%), while non-English data was split at a ratio of 75\% for transcription and 25\% for translation to English.
Finally, for the ASR \& ST ($X \leftrightarrow \text{en}$) configuration, this 75:25 ratio of transcription to translation was applied uniformly across all languages.

\textbf{Speech LLM Training Data:} For the subsequent phase, where the trainable adaptor maps the extracted speech representations into the input space of the frozen LLM, we curated a multi-task dataset totaling approximately 6.2k hours.
Specifically, this dataset encompasses transcription, translation, intent classification, and emotion recognition, thereby equipping the integrated Speech LLM with diverse generative capabilities.
The core transcription task covered four target languages: English (en), Japanese (ja), Chinese (zh), and German (de).
The translation tasks includ bidirectional mapping between English and three languages (ja, zh, and de), as well as unidirectional translation from English to four languages unseen during the pre-training phase: Persian (fa), Indonesian (id), Swedish (sv), and Turkish (tr).
For the core transcription and translation tasks, we combined several established datasets: VoxPopuli~\cite{wang-etal-2021-voxpopuli}, FLEURS~\cite{fleurs}, AISHELL~\cite{aichell1}, JSUT~\cite{jsut:2017}, CoVoST2~\cite{wang2020covost}, and SpeechBSD~\cite{shimizu-etal-2023-towards}.
To ensure robustness, we also blended in 300 hours of transcription data and 200 hours of translation data, randomly sampled from our pre-training corpora across all four languages (English, Japanese, German, and Chinese).
For spoken language understanding and paralinguistic tasks, we utilized SLURP~\cite{bastianelli-etal-2020-slurp} (English) and the German subset of Speech-MASSIVE~\cite{lee2024speechmassivemultilingualspeechdataset} for intent classification, alongside MELD~\cite{poria-etal-2019-meld} (English) for emotion recognition.
To prevent bias toward the dominant transcription and translation tasks, we mitigated data imbalance through strategic oversampling (e.g., upsampling the MELD dataset by a factor of 5).

\textbf{Implementation Details:} Our training pipeline consisted of two distinct stages: base model pre-training and Speech LLM training.
For the base model pre-training, the model was trained for 3 epochs with a global batch size of 512.
Referencing the approach in OWSM v3.1~\cite{owsm-v3.1}, we employed a piecewise-linear warmup, increasing the learning rate to $5 \times 10^{-5}$ over the initial 15k steps and to its peak over the subsequent 15k steps.
The peak learning rate was set to $2 \times 10^{-4}$, following the original Whisper settings~\cite{Radford2023whisper}.
After this 30k-step warmup period, the learning rate followed a cosine decay schedule.
This pre-training was conducted on 16 NVIDIA H100 GPUs.

For the subsequent Speech LLM training stage, we optimized only the adaptor for 25k steps with a global batch size of 512.
The peak learning rate was set to $1 \times 10^{-4}$ with a 500-step linear warmup, followed by a cosine decay schedule.
This training stage was conducted on 8 NVIDIA H100 GPUs.

\subsection{Model Configurations}
To implement our Speech LLM, we utilize the encoder portion of the Whisper medium architecture as our speech representation extractor.
For the core language modeling component, we employ two variants of the Llama 3.2 family: Llama-3.2-1B-Instruct and Llama-3.2-3B-Instruct.

To bridge the modality gap between continuous audio signals and the discrete text space, we designed a lightweight trainable adaptor comprising a two-layer convolutional neural network (CNN) followed by a linear projection layer. 
The CNN layers effectively downsample the temporal resolution of the speech sequence, thereby reducing the computational burden on the LLM, while the linear layer projects the compressed acoustic features into the exact embedding dimensionality of the LLM.

\subsection{Evaluation Setup and Metrics}
\textbf{Tasks and Metrics:} We comprehensively evaluate the integrated Speech LLM across four downstream tasks: ASR, ST, intent classification, and emotion recognition.
For the ASR task, we measure the Word Error Rate (WER) for English and German, and the Character Error Rate (CER) for Japanese and Chinese on the FLEURS benchmark.
For the ST task, we evaluate the generation quality by reporting BLEU scores on both the FLEURS and CoVoST2, covering the same bidirectional and unidirectional language pairs used during training.
Furthermore, we calculate the classification accuracy (Acc) for Intent Classification on SLURP and Speech-MASSIVE, as well as for emotion recognition on MELD.
Evaluating this set of tasks allows us to systematically assess the generalizability of the pre-training configurations across a wide range of downstream capabilities.
To ensure the robustness of our results, all reported metrics for each task represent the average values obtained from three independent inference runs.

\textbf{Evaluated Configurations:} In our primary experimental setting, the parameters of both the speech encoder and the LLM were kept strictly frozen during the training phase.
This strict constraint ensures that any variation in downstream performance is solely attributable to the intrinsic quality of the encoder's pre-trained representations. 
Because the adaptor is the only trainable component, this setup allows us to purely evaluate how effectively each pre-training configuration extracts representations that align with the LLM's text-based input space.
Furthermore, to provide a comprehensive analysis, we explored a supplementary configuration where the speech encoder was jointly updated with the adaptor during the Speech LLM training phase.
Since some existing Speech LLMs~\cite{chu2023qwenaudioadvancinguniversalaudio,xu2025qwen25omnitechnicalreport} employ joint training of the encoder and the LLM, this experiment allows us to assess whether the benefits of our pre-training tasks persist when the encoder is further adapted during the final integration stage.

\subsection{Results}
Table~\ref{tab:main_results} summarizes the downstream performance on the ASR and ST tasks. 
We first observe that translation-enhanced pre-training substantially outperforms the ASR-only baseline across both the 1B and 3B LLM scales, indicating that the benefits of incorporating speech translation into pre-training are consistent across different language model capacities.

Specifically, the inclusion of ST ($X \rightarrow \text{en}$) during pre-training consistently improves downstream translation performance over the ASR-only setting.
More importantly, this configuration also markedly reduces ASR error rates for non-English languages.
For instance, in the 1B model, the CER of Japanese ASR drops from 29.2 to 21.1.
While English ASR also shows modest gains, the impact is most evident in the non-English languages that were used as audio inputs for the translation task.
This indicates that requiring the encoder to process non-English audio for translation effectively refines its ability to extract acoustic representations for those specific languages.

Furthermore, introducing English-to-non-English translation ($\text{en} \rightarrow X$) to form a bidirectional configuration ($X \leftrightarrow \text{en}$) drastically improves $\text{en} \rightarrow X$ ST performance.
Notably, this improvement extends even to target languages (fa, id, sv, tr) that were unseen during the Seq2Seq pre-training phase.
These results demonstrate that incorporating translation into pre-training has a profound effect on unlocking the frozen LLM's inherent multilingual capabilities, extending well beyond the specific language pairs seen during the pre-training phase.

\begin{table}[t]
\centering
\caption{Downstream performance on classification tasks (Accuracy $\uparrow$) for 3B model. ``Intent'' involves understanding user commands, evaluated on SLURP and Speech-massive. ``Emotion'' is evaluated on MELD.}
\label{tab:classification_results}
\vspace{-3mm}
\begin{tabular}{l ccc}
\toprule
& \multicolumn{2}{c}{\textbf{Intent}} & \textbf{Emotion} \\
\cmidrule(lr){2-3} \cmidrule(lr){4-4}
\textbf{Pre-training Task Mixture} & \textbf{en} & \textbf{de} & \textbf{en} \\
\midrule
ASR Only                           & 57.3 & 57.9 & 49.2 \\
ASR \& ST (X $\rightarrow$ en)     & 58.5 & 62.1 & \bf{50.3} \\
ASR \& ST (X $\leftrightarrow$ en) & \bf{64.5} & \bf{66.3} & 49.5 \\
\bottomrule
\end{tabular}
\vspace{-4mm}
\end{table}
\begin{table}[t]
\centering
\caption{Downstream translation performance (Avg. BLEU $\uparrow$) for Llama-3.2-3B with an \textbf{unfrozen} encoder. Left column shows \textbf{FLEURS} scores for X $\rightarrow$ en; right column shows \textbf{CoVoST 2 / FLEURS} scores for en $\rightarrow$ X.}
\label{tab:unfrozen_st_results}
\vspace{-3mm}
\resizebox{\columnwidth}{!}{
\begin{tabular}{l cc}
\toprule
& \textbf{ST} (X $\rightarrow$ en) & \textbf{ST} (en $\rightarrow$ X) \\
\textbf{Pre-training Task Mixture} & (Avg. BLEU $\uparrow$) & (Avg. BLEU $\uparrow$) \\
\midrule
ASR Only                           & 20.1 & 20.4/21.0 \\
ASR \& ST (X $\rightarrow$ en)     & 22.1 & 21.1/21.5 \\
ASR \& ST (X $\leftrightarrow$ en) & \bf{22.9} & \bf{21.9}/\bf{22.5} \\
\bottomrule
\end{tabular}
}
\vspace{-4mm}
\end{table}

Next, to assess the impact of our pre-training objectives on spoken language understanding, we evaluate the 3B model on intent (SLURP, Speech-MASSIVE) and emotion (MELD) classification (Table~\ref{tab:classification_results}).
These tasks go beyond mere phonetic matching, requiring the model to extract underlying user goals and paralinguistic cues.
Crucially, the performance gains directly mirror the source languages used in translation pre-training.
In the unidirectional ($X \rightarrow \text{en}$) setup, German intent accuracy improves notably (57.9 to 62.1), whereas English sees only marginal gains (57.3 to 58.5) as its English input was limited to transcription during pre-training
However, the bidirectional ($X \leftrightarrow \text{en}$) objective significantly increases English intent accuracy to 64.5 and further boosts German to 66.3.
In contrast, performance on emotion recognition remains largely unaffected by the choice of pre-training configuration. 
Because this task relies on fine-grained acoustic information rather than pure semantics, translation objectives do not yield gains. 
Importantly, the lack of degradation confirms that semantic abstraction through translation does not harm acoustic-dependent tasks. 
Thus, a symmetric translation objective enhances semantic understanding while safely preserving essential acoustic cues.

\textbf{Impact of Unfreezing the Encoder:} 
We also evaluate a setting where the speech encoder is unfrozen, similar to the training pipelines used in several existing Speech LLMs.
As shown in Table~\ref{tab:unfrozen_st_results}, while unfreezing generally improves performance, the bidirectional configuration ($X \leftrightarrow \text{en}$) maintains its clear advantage over the baseline and unidirectional models. 
This result confirms that the benefits of bidirectional pre-training persist even after joint optimization. 
It suggests that our symmetric objective provides a superior representation that cannot be replicated by simply unfreezing the encoder during downstream adaptation.

\section{Conclusion}
In this work, we demonstrated that symmetric, bidirectional translation ($X \leftrightarrow \text{en}$) is a highly effective pre-training objective for Speech LLMs. 
Compared to standard transcription or unidirectional baselines, this approach significantly improves downstream performance across both speech translation and classification tasks.
Intent classification accuracy improves for languages explicitly used as translation sources, while performance on acoustic-dependent tasks like emotion recognition remains undegraded.
Thus, bidirectional pre-training provides an optimal foundation by deepening semantic abstraction while safely preserving essential acoustic cues.

While massive scaling might eventually lead to semantic abstraction, our results show that incorporating bidirectional translation provides a more direct and reliable path to language-agnostic representations without degrading ASR and acoustic-sensitive tasks. 
This suggests that such a symmetric objective should be considered a standard pre-training practice for robust Speech LLMs.

\newpage
\section{Generative AI Use Disclosure}
Generative AI tools (Gemini and ChatGPT) were used for language editing and improving the phrasing of this manuscript.

\bibliographystyle{IEEEtran}
\bibliography{references}

@inproceedings{Radford2023whisper,
    author = {Radford, Alec and Kim, Jong Wook and Xu, Tao and Brockman, Greg and McLeavey, Christine and Sutskever, Ilya},
    title = {{Robust speech recognition via large-scale weak supervision}},
    year = {2023},
    booktitle = {Proceedings of the 40th International Conference on Machine Learning},
}

@inproceedings{tang2024salmonn,
  title={{SALMONN: Towards Generic Hearing Abilities for Large Language Models}},
  author={Changli Tang and Wenyi Yu and Guangzhi Sun and Xianzhao Chen and Tian Tan and Wei Li and Lu Lu and Zejun MA and Chao Zhang},
  booktitle={Proceedings of the Twelfth International Conference on Learning Representations},
  year={2024},
}

@misc{rubenstein2023audiopalm,
      title={{AudioPaLM: A Large Language Model That Can Speak and Listen}}, 
      author={Paul K. Rubenstein and Chulayuth Asawaroengchai and Duc Dung Nguyen and Ankur Bapna and Zalán Borsos and others},
      year={2023},
      note={arXiv:2306.12925},
}

@misc{zhang2023speechgpt,
      title={{SpeechGPT: Empowering Large Language Models with Intrinsic Cross-Modal Conversational Abilities}}, 
      author={Dong Zhang and Shimin Li and Xin Zhang and Jun Zhan and Pengyu Wang and Yaqian Zhou and Xipeng Qiu},
      year={2023},
      eprint={2305.11000},
      archivePrefix={arXiv},
      primaryClass={cs.CL},
      note={arXiv:2305.11000}
}

@inproceedings{schneider19_interspeech,
  author={Steffen Schneider and Alexei Baevski and Ronan Collobert and Michael Auli},
  title={{wav2vec: Unsupervised Pre-Training for Speech Recognition}},
  year=2019,
  booktitle={Proceedings of Interspeech 2019},
  pages={3465--3469},
}

@inproceedings{choi2024_interspeech,
  author={Kwanghee Choi and Ankita Pasad and Tomohiko Nakamura and Satoru Fukayama and Karen Livescu and Shinji Watanabe},
  title={{Self-Supervised Speech Representations are More Phonetic than Semantic}},
  year=2024,
  booktitle={Proceedings of Interspeech 2024},
  pages={4578--4582},
}

@INPROCEEDINGS{fathullah2023prompting,
    title={{Prompting Large Language Models with Speech Recognition Abilities}}, 
    author={Fathullah, Yassir and Wu, Chunyang and Lakomkin, Egor and Jia, Junteng and Shangguan, Yuan and Li, Ke and Guo, Jinxi and Xiong, Wenhan and Mahadeokar, Jay and Kalinli, Ozlem and Fuegen, Christian and Seltzer, Mike},
    booktitle={Proceedings of the 2024 IEEE International Conference on Acoustics, Speech and Signal Processing}, 
    year={2024},
    pages={13351--13355},
  }

@INPROCEEDINGS{Lakomkin2024,
  author={Lakomkin, Egor and Wu, Chunyang and Fathullah, Yassir and Kalinli, Ozlem and Seltzer, Michael L. and Fuegen, Christian},
  booktitle={Proceedings of the 2024 IEEE International Conference on Acoustics, Speech and Signal Processing}, 
  title={{End-to-End Speech Recognition Contextualization with Large Language Models}}, 
  year={2024},
  pages={12406-12410},
}

@article{ma2024embarrassingly,
  title={{An Embarrassingly Simple Approach for LLM with Strong ASR Capacity}},
  author={Ma, Ziyang and Yang, Guanrou and Yang, Yifan and Gao, Zhifu and Wang, Jiaming and others},
  journal={arXiv:2402.08846},
  year={2024}
}

@inproceedings{hono-etal-2024-integrating,
    title = {{Integrating Pre-Trained Speech and Language Models for End-to-End Speech Recognition}},
    author = "Hono, Yukiya  and
      Mitsuda, Koh  and
      Zhao, Tianyu  and
      Mitsui, Kentaro  and
      Wakatsuki, Toshiaki  and
      Sawada, Kei",
    booktitle = "Findings of the Association for Computational Linguistics",
    year = "2024",
    pages = "13289--13305",
}

@inproceedings{comfomer,
  author = {Anmol Gulati and James Qin and Chung-Cheng Chiu and Niki Parmar and Yu Zhang and Jiahui Yu and Wei Han and Shibo Wang and Zhengdong Zhang and Yonghui Wu and Ruoming Pang},
  booktitle = {Proceedings of the Interspeech 2020},
  title = {{Conformer: Convolution-augmented Transformer for Speech Recognition}},
  year = {2020},
  pages = {5036--5040},
}

@article{10.1109/TASLP.2021.3122291, 
   author = {Hsu, Wei-Ning and Bolte, Benjamin and Tsai, Yao-Hung Hubert and Lakhotia, Kushal and Salakhutdinov, Ruslan and Mohamed, Abdelrahman}, 
   title = {{HuBERT: Self-Supervised Speech Representation Learning by Masked Prediction of Hidden Units}}, 
   year = {2021}, 
   journal = {IEEE/ACM Transactions on Audio, Speech and Language Processing}, 
   pages = {3451--3460}, 
   numpages = {10} 
}

@INPROCEEDINGS{peng2023asru,
  author={Peng, Yifan and Tian, Jinchuan and Yan, Brian and Berrebbi, Dan and Chang, Xuankai and Li, Xinjian and Shi, Jiatong and Arora, Siddhant and Chen, William and Sharma, Roshan and Zhang, Wangyou and Sudo, Yui and Shakeel, Muhammad and Jung, Jee-Weon and Maiti, Soumi and Watanabe, Shinji},
  booktitle={Proceedings of the 2023 IEEE Automatic Speech Recognition and Understanding Workshop}, 
  title={{Reproducing Whisper-Style Training Using An Open-Source Toolkit And Publicly Available Data}}, 
  year={2023},
}

@INPROCEEDINGS{wang2023asru,
  author={Wang, Mingqiu and Han, Wei and Shafran, Izhak and Wu, Zelin and Chiu, Chung-Cheng and Cao, Yuan and Chen, Nanxin and Zhang, Yu and Soltau, Hagen and Rubenstein, Paul K. and Zilka, Lukas and Yu, Dian and Pundak, Golan and Siddhartha, Nikhil and Schalkwyk, Johan and Wu, Yonghui},
  booktitle={Proceedings of the 2023 IEEE Automatic Speech Recognition and Understanding Workshop}, 
  title={{SLM: Bridge the Thin Gap Between Speech and Text Foundation Models}}, 
  year={2023},
}

@inproceedings{fang2025llamaomni,
  title={{LLaMA-Omni: Seamless Speech Interaction with Large Language Models}},
  author={Qingkai Fang and Shoutao Guo and Yan Zhou and Zhengrui Ma and Shaolei Zhang and Yang Feng},
  booktitle={Proceedings of the Thirteenth International Conference on Learning Representations},
  year={2025},
}

@inproceedings{zhang-etal-2025-soundwave,
    title = {{Soundwave: Less is More for Speech-Text Alignment in LLMs}},
    author = "Zhang, Yuhao  and
      Liu, Zhiheng  and
      Bu, Fan  and
      Zhang, Ruiyu  and
      Wang, Benyou  and
      Li, Haizhou",
    booktitle = "Proceedings of the 63rd Annual Meeting of the Association for Computational Linguistics",
    year = "2025",
    pages = "18718--18738",
}

@misc{chu2024qwen2audiotechnicalreport,
      title={{Qwen2-Audio Technical Report}}, 
      author={Yunfei Chu and Jin Xu and Qian Yang and Haojie Wei and Xipin Wei and Zhifang Guo and Yichong Leng and Yuanjun Lv and Jinzheng He and Junyang Lin and Chang Zhou and Jingren Zhou},
      year={2024},
      eprint={2407.10759},
      archivePrefix={arXiv},
      primaryClass={eess.AS},
      url={https://arxiv.org/abs/2407.10759}, 
      note = "arXiv:2407.10759",
}

@misc{xu2025qwen25omnitechnicalreport,
      title={{Qwen2.5-Omni Technical Report}}, 
      author={Jin Xu and Zhifang Guo and Jinzheng He and Hangrui Hu and Ting He and Shuai Bai and Keqin Chen and Jialin Wang and Yang Fan and Kai Dang and Bin Zhang and Xiong Wang and Yunfei Chu and Junyang Lin},
      year={2025},
      note = "arXiv:2503.20215",
}

@inproceedings{owsm-v4,
  title={{OWSM v4: Improving Open Whisper-Style Speech Models via Data Scaling and Cleaning}},
  author={Yifan Peng and Shakeel Muhammad and Yui Sudo and William Chen and Jinchuan Tian and Chyi-Jiunn Lin and Shinji Watanabe},
  booktitle={Proceedings of the Interspeech 2025},
  year={2025},
}

@inproceedings{owsm-v3.1,
  title={{OWSM v3.1: Better and Faster Open Whisper-Style Speech Models based on E-Branchformer}},
  author={Yifan Peng and Jinchuan Tian and William Chen and Siddhant Arora and Brian Yan and Yui Sudo and Muhammad Shakeel and Kwanghee Choi and Jiatong Shi and Xuankai Chang and Jee-weon Jung and Shinji Watanabe},
  booktitle={Proceedings of the Interspeech 2024},
  year={2024},
}

@misc{qwen2025qwen25technicalreport,
  title={{Qwen2.5 Technical Report}}, 
  author={Qwen and : and An Yang and Baosong Yang and Beichen Zhang and Binyuan Hui and Bo Zheng and Bowen Yu and Chengyuan Li and Dayiheng Liu and Fei Huang and Haoran Wei and others},
  year={2025},
  note={arXiv:2412.15115},
}

@inproceedings{wang2020covost,
    title={{CoVoST 2: A Massively Multilingual Speech-to-Text Translation Corpus}},
    author={Changhan Wang and Anne Wu and Juan Pino},
    booktitle={Proceedings of the Interspeech 2021},
    year={2021},
}

@inproceedings{reazonspeech,
  title="{ReazonSpeech: A free and massive corpus for Japanese ASR}",
  author={Yin, Yue and Mori, Daijiro and Fujimoto, Seiji},
  booktitle={Proceedings of the 29th Annual Meeting of the Association for Natural Language Processing (Domestic Conference)},
  pages = {1134--1139},
  year={2023}
}

@inproceedings{librispeech,
  author={Panayotov, Vassil and Chen, Guoguo and Povey, Daniel and Khudanpur, Sanjeev},
  booktitle={Proceedings of the 2015 IEEE International Conference on Acoustics, Speech and Signal Processing}, 
  title={{Librispeech: An ASR corpus based on public domain audio books}}, 
  year={2015},
  pages={5206-5210},
}

@inproceedings{Pratap2020MLSAL,
  title={{MLS: A Large-Scale Multilingual Dataset for Speech Research}},
  author={Vineel Pratap and Qiantong Xu and Anuroop Sriram and Gabriel Synnaeve and Ronan Collobert},
  booktitle={Proceedings of the Interspeech 2020},
  year={2020},
}

@inproceedings{zhang2022wenetspeech,
  title={{Wenetspeech: A 10000+ hours multi-domain mandarin corpus for speech recognition}},
  author={Zhang, Binbin and Lv, Hang and Guo, Pengcheng and Shao, Qijie and Yang, Chao and Xie, Lei and Xu, Xin and Bu, Hui and Chen, Xiaoyu and Zeng, Chenchen and others},
  booktitle={Proceedings of the 2022 IEEE International Conference on Acoustics, Speech and Signal Processing},
  pages={6182--6186},
  year={2022},
}

@inproceedings{commonvoice:2020,
  author = {Ardila, R. and Branson, M. and Davis, K. and Henretty, M. and Kohler, M. and Meyer, J. and Morais, R. and Saunders, L. and Tyers, F. M. and Weber, G.},
  title = {{Common Voice: A Massively-Multilingual Speech Corpus}},
  booktitle = {Proceedings of the 12th Conference on Language Resources and Evaluation},
  pages = {4211--4215},
  year = 2020
}

@article{jsut:2017,
  author={Ryosuke Sonobe and Shinnosuke Takamichi and Hiroshi Saruwatari},
  title={{JSUT} corpus: free large-scale Japanese speech corpus for end-to-end speech synthesis},
  journal={arXiv:1711.00354},
  year={2017},
}

@inproceedings{shimizu-etal-2023-towards,
    title = "{Towards Speech Dialogue Translation Mediating Speakers of Different Languages}",
    author = "Shimizu, Shuichiro  and
      Chu, Chenhui  and
      Li, Sheng  and
      Kurohashi, Sadao",
    booktitle = "Findings of the Association for Computational Linguistics: ACL 2023",
    year = "2023",
    pages = "1122--1134",
}

@inproceedings{aichell1,
  author={Bu, Hui and Du, Jiayu and Na, Xingyu and Wu, Bengu and Zheng, Hao},
  booktitle={Proceedings of the 20th Conference of the Oriental Chapter of the International Coordinating Committee on Speech Databases and Speech I/O Systems and Assessment}, 
  title={{AISHELL-1: An open-source Mandarin speech corpus and a speech recognition baseline}}, 
  year={2017},
}

@inproceedings{fleurs,
  author={Conneau, Alexis and Ma, Min and Khanuja, Simran and Zhang, Yu and Axelrod, Vera and Dalmia, Siddharth and Riesa, Jason and Rivera, Clara and Bapna, Ankur},
  booktitle={Proceedings of the 2023 IEEE Spoken Language Technology Workshop}, 
  title={{FLEURS: FEW-Shot Learning Evaluation of Universal Representations of Speech}}, 
  year={2023},
  pages={798--805},
}

@inproceedings{wang-etal-2021-voxpopuli,
    title = "{VoxPopuli: A Large-Scale Multilingual Speech Corpus for Representation Learning, Semi-Supervised Learning and Interpretation}",
    author = "Wang, Changhan  and
      Riviere, Morgane  and
      Lee, Ann  and
      Wu, Anne  and
      Talnikar, Chaitanya  and
      Haziza, Daniel  and
      Williamson, Mary  and
      Pino, Juan  and
      Dupoux, Emmanuel",
    booktitle = "Proceedings of the 59th Annual Meeting of the Association for Computational Linguistics and the 11th International Joint Conference on Natural Language Processing",
    year = "2021",
    pages = "993--1003",
}

@inproceedings{bastianelli-etal-2020-slurp,
    title = "{SLURP: A Spoken Language Understanding Resource Package}",
    author = "Bastianelli, Emanuele  and
      Vanzo, Andrea  and
      Swietojanski, Pawel  and
      Rieser, Verena",
    booktitle = "Proceedings of the 2020 Conference on Empirical Methods in Natural Language Processing",
    year = "2020",
    pages = "7252--7262",
}

@inproceedings{poria-etal-2019-meld,
    title = "{MELD: A Multimodal Multi-Party Dataset for Emotion Recognition in Conversations}",
    author = "Poria, Soujanya  and
      Hazarika, Devamanyu  and
      Majumder, Navonil  and
      Naik, Gautam  and
      Cambria, Erik  and
      Mihalcea, Rada",
    booktitle = "Proceedings of the 57th Annual Meeting of the Association for Computational Linguistics",
    year = "2019",
    pages = "527--536",
}

@inproceedings{lee2024speechmassivemultilingualspeechdataset,
      title={{Speech-MASSIVE: A Multilingual Speech Dataset for SLU and Beyond}}, 
      author={Beomseok Lee and Ioan Calapodescu and Marco Gaido and Matteo Negri and Laurent Besacier},
      year={2024},
      booktitle={Proceedings of the Interspeech 2024},
}

@inproceedings{ma-etal-2025-cross,
    title = "{Cross-Lingual Transfer Learning for Speech Translation}",
    author = "Ma, Rao  and
      Qian, Mengjie  and
      Fathullah, Yassir  and
      Tang, Siyuan  and
      Gales, Mark  and
      Knill, Kate",
    booktitle = "Proceedings of the 2025 Conference of the Nations of the Americas Chapter of the Association for Computational Linguistics: Human Language Technologies",
    year = "2025",
    pages = "33--43",
}

@inproceedings{liu2026ttatranscribetranslatealignment,
      title={{TTA: Transcribe, Translate and Alignment for Cross-lingual Speech Representation}}, 
      author={Wei Liu and Jiahong Li and Yiwen Shao and Dong Yu},
      year={2026},
      booktitle = "Proceedings of the 2026 IEEE International Conference on Acoustics, Speech and Signal Processing",
}

@misc{xu2025qwen3omnitechnicalreport,
      title={{Qwen3-Omni Technical Report}}, 
      author={Jin Xu and Zhifang Guo and Hangrui Hu and Yunfei Chu and Xiong Wang and Jinzheng He and Yuxuan Wang and Xian Shi and Ting He and Xinfa Zhu and Yuanjun Lv and Yongqi Wang and Dake Guo and He Wang and Linhan Ma and Pei Zhang and Xinyu Zhang and Hongkun Hao and Zishan Guo and Baosong Yang and Bin Zhang and Ziyang Ma and Xipin Wei and Shuai Bai and Keqin Chen and Xuejing Liu and Peng Wang and Mingkun Yang and Dayiheng Liu and Xingzhang Ren and Bo Zheng and Rui Men and Fan Zhou and Bowen Yu and Jianxin Yang and Le Yu and Jingren Zhou and Junyang Lin},
      year={2025},
      note = "arXiv:2509.17765",
}

@inproceedings{hu-etal-2024-wavllm,
    title = "{WavLLM: Towards Robust and Adaptive Speech Large Language Model}",
    author = "Hu, Shujie  and
      Zhou, Long  and
      Liu, Shujie  and
      Chen, Sanyuan  and
      Meng, Lingwei  and
      Hao, Hongkun  and
      Pan, Jing  and
      Liu, Xunying  and
      Li, Jinyu  and
      Sivasankaran, Sunit  and
      Liu, Linquan  and
      Wei, Furu",
    booktitle = "Findings of the Association for Computational Linguistics: EMNLP 2024",
    year = "2024",
    pages = "4552--4572",
}

@misc{chu2023qwenaudioadvancinguniversalaudio,
      title={{Qwen-Audio: Advancing Universal Audio Understanding via Unified Large-Scale Audio-Language Models}}, 
      author={Yunfei Chu and Jin Xu and Xiaohuan Zhou and Qian Yang and Shiliang Zhang and Zhijie Yan and Chang Zhou and Jingren Zhou},
      year={2023},
      note = "arXiv:2311.07919",
}

@misc{fang2025llamaomni2llmbasedrealtimespoken,
      title={LLaMA-Omni2: LLM-based Real-time Spoken Chatbot with Autoregressive Streaming Speech Synthesis}, 
      author={Qingkai Fang and Yan Zhou and Shoutao Guo and Shaolei Zhang and Yang Feng},
      year={2025},
      eprint={2505.02625},
      archivePrefix={arXiv},
      primaryClass={cs.CL},
      url={https://arxiv.org/abs/2505.02625}, 
}

\end{document}